\documentclass[reprint,showpacs,amsmath,amssymb,pra]{revtex4-1}
\usepackage{graphicx}
\usepackage{color}
\usepackage{dcolumn}
\usepackage{bm}
\usepackage{tabularx}

\begin{document}

\title{Disorder-induced light trapping enhanced by pulse collisions in one-dimensional nonlinear photonic crystals}

\author{Denis V. Novitsky}
\email{Corresponding author: dvnovitsky@tut.by} \affiliation{B. I.
Stepanov Institute of Physics, National Academy of Sciences of
Belarus, Nezavisimosti Avenue 68, BY-220072 Minsk, Belarus}

\begin{abstract}
We use numerical simulations to study interaction of co- and
counter-propagating pulses in disordered multilayers with
noninstantaneous Kerr nonlinearity. We propose a statistical
argument for existence of the disorder-induced trapping which
implies the dramatic rise of the probability of realization with low
output energy in the structure with a certain level of disorder.
This effect is much more pronounced in the case of two interacting
pulses than in the single-pulse regime and does not occur in the
strictly ordered system at the same intensity of the pulses.
Therefore it cannot be explained simply as a result of increase in
strength of nonlinear light-matter interaction.
\end{abstract}


\maketitle

\section{Introduction}

Since Philip W. Anderson's breakthrough paper \cite{Anderson}, study
of localization and other matter-wave effects in solid-state
disordered systems has become a broad and fruitful field of
research. Moreover, the notion of Anderson localization stimulated
research of wave phenomena in other contexts, including classical
wave dynamics in disordered media and the connections with
mesoscopic physics \cite{Lagendijk, Abrahams, Sheng, Tiggelen}. In
optics, this interest has led to the experimental observation of the
Anderson localization of light in 1990s and 2000s \cite{Wiersma,
Chabanov, Storzer}. Discussion of subsequent progress in disordered
optics and photonics can be found in recent reviews
\cite{WiersmaRev, Segev}.

In this paper, we deal with some aspects of nonlinear optics of
disordered photonic structures. For detailed discussion of
short-pulse effects (including tail dynamics \cite{Conti2007},
localization suppression \cite{Pikovsky, Flach}, localized solitons
formation \cite{Conti2012, Folli2011}, etc.) in nonlinear disordered
systems, see the introduction to my previous paper
\cite{Novitsky2014} and references therein. Here we restrict
ourselves to referring only to a few recent advances reported in
literature. Among them are the observation of the reciprocity
breaking effect in nonlinear random medium \cite{Muskens}, the
parametric amplification of light localization in the random medium
with quadratic nonlinearity \cite{Folli2013}, self-trapping of light
in nonlinear waveguide array with coupling disorder \cite{Naether1},
nonreciprocal localization in disordered multilayers with
magneto-optical materials \cite{Bliokh}, wave packet spreading in 1D
and 2D photonic lattices \cite{Naether2}, control of energy transfer
in disordered laser resonators \cite{Leonetti}, etc.

This paper can be viewed as a continuation of the previous work
\cite{Novitsky2014} devoted to propagation and self-trapping of
ultrashort pulses in disordered one-dimensional photonic crystals
with instantaneous and relaxing nonlinearities. Here we consider the
collisions of pulses in such structures and search for the
possibility of light trapping which cannot be reached in ordered
system with the same parameters. This trapping is fundamentally
different from the self-trapping effect in the perfect nonlinear
photonic crystals \cite{Novitsky2010} which is destroyed by
introduction of disorder. As previously, we consider the regime of
strong disorder and strong nonlinearity. We have studied earlier the
interaction of co- and counter-propagating pulses in perfect
photonic crystals with relaxing nonlinearity \cite{Novitsky2013} and
in dense two-level media \cite{Novit2011, Novit2012, Novit2014}. As
far as we know, the influence of disorder on such interaction was
not considered in scientific literature yet. The present study makes
up for this deficiency.

The paper is structured as follows. In Section \ref{state}, we give
the main equations and briefly discuss the numerical method and the
parameters adopted. Sections \ref{coprop} and \ref{counterprop} are
dedicated to the analysis of results obtained for co- and
counter-propagating pulses, respectively. The paper is completed
with the short Conclusion.

\section{\label{state}Problem statement}

Let us consider the one-dimensional photonic crystal, i.e. a
multilayer structure consisting of two different materials --
alternating layers denoted with letters $a$ and $b$. Light is
assumed to propagate along the $z$-axis which is perpendicular to
the layers interfaces. The results reported here are based on
numerical solution of the one-dimensional wave equation,
\begin{eqnarray}
\frac{\partial^2 E}{\partial z^2}&-&\frac{1}{c^2} \frac{\partial^2
(n^2 E)}{\partial t^2} = 0, \label{Max}
\end{eqnarray}
where $E$ is the electric field strength, $n$ is the medium
refractive index which, generally, is a function light intensity
$I=|E|^2$,
\begin{eqnarray}
n=n^0(z)+\delta n (I, t, z). \label{refr}
\end{eqnarray}
Here $n^0(z)$ is a linear part of refractive index changing
periodically along the structure. Since we deal with
noninstantaneous nonlinearity, the nonlinear contribution $\delta n$
must take into account the relaxation process which, for
definiteness, will be described by the Debye model \cite{Akhm},
\begin{eqnarray}
t_{nl} \frac{d \delta n}{d t}+ \delta n=n_2 I, \label{relax}
\end{eqnarray}
where $n_2$ is the cubic (Kerr) nonlinear coefficient, and $t_{nl}$
is the relaxation time. For the disordered periodic structure, we
assume the random variations of thicknesses of layers $a$ and $b$ as
follows,
\begin{eqnarray}
d_{a,b}=d_{a,b}^0+\Delta d (\xi-1/2), \label{rand}
\end{eqnarray}
where $d_{a,b}^0$ are the mean values of thicknesses, $\Delta d$ is
the amplitude of disorder, and $\xi$ is the random quantity
uniformly distributed in the range $[0, 1]$.

We solve numerically Eqs. \ref{Max}-\ref{rand} using the method
developed in the previous publications \cite{Novitsky2010,
Novitsky2014}. As previously, we do not mean any specific materials,
since our aim is to study the qualitative and general aspects of
light interactions with periodic disordered structures. Therefore,
for our calculations, we adopt the parameters of the model from Ref.
\cite{Novitsky2014}: $d_a^0=0.4$ and $d_b^0=0.24$ $\mu$m, $n^0_a=2$
and $n^0_b=1.5$. The envelope of the pulse at the input of the
photonic structure is supposed to have the Gaussian shape, $A(t)=A_0
\exp(-t^2/2t_p^2)$, where $t_p$ is the pulse duration, and $A_0$ is
the amplitude of the electric field. Further we assume $t_p=50$ fs
and the central wavelength $\lambda_c=1.064$ $\mu$m, so that the
carrier frequency lies just outside the band gap of the perfect
multilayer \cite{Novitsky2014}. Finally, we restrict ourselves to
the structure with nonlinear $b$ layers only. This is justified,
because light concentrates in these layers when, as in our case, we
deal with the high-frequency edge of the band gap \cite{Bertolotti}.
The strength of nonlinearity ($n_2 I_0=n_2 |A_0|^2 \sim 0.01$) is
taken to be large enough to strongly influence the pulse
characteristics. This allows to consider comparatively short
systems, namely $N=50$ periods in our calculations. Construction of
such photonic crystals seems to be quite feasible for modern
technology. Though we do not mean any specific materials, linear
layers may be formed by glass, while for nonlinear layers one can
use polymer materials possessing high nonlinearity and fast
relaxation \cite{Meng}. However, as far as we know, such photonic
crystals possessing relaxing nonlinearity and disorder
simultaneously were not realized experimentally yet. Therefore, our
study can be considered as a proposal for building such new optical
systems as well.

Thus, we consider the interplay of strong disorder and strong
nonlinearity. Generally, this interplay can be studied on the short
timescale (pulse shape transformation) and at long times (pulse tail
transformation as an evidence for the Anderson localization) as was
done in the previous work \cite{Novitsky2014}. In this paper, we
deal with the collisions of pulses in the disordered photonic
crystals. Since the behavior of the tail and the Anderson
localization seem to be insensitive to the number of pulses, we will
focus on the shape transformations of the colliding pulses and, in
particular, on the possibility to induce light trapping by using the
collisions of co- and counter-propagating pulses.

\section{\label{coprop}Co-propagating pulses}

\begin{figure}[t!]
\includegraphics[scale=0.9, clip=]{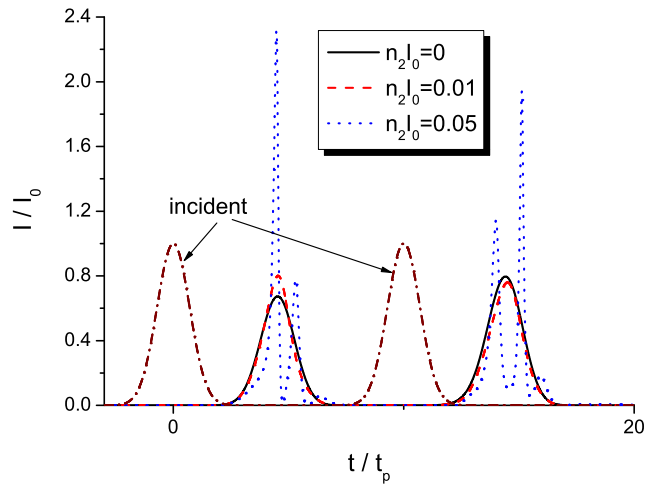}
\centering \caption{\label{fig1} (Color online) The profiles of
co-propagating pulses transmitted through the perfect (ordered)
photonic crystal with and without nonlinearity.}
\end{figure}

\begin{figure}[t!]
\includegraphics[scale=0.9, clip=]{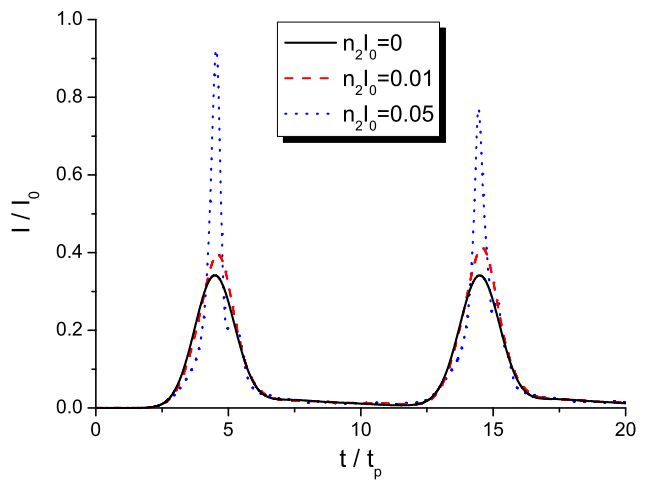}
\centering \caption{\label{fig2} (Color online) The same as in Fig.
\ref{fig1} but for the disordered structure with $\Delta d=0.05$
$\mu$m. The curves are averaged over $50$ realizations.}
\end{figure}

\begin{figure}[t!]
\includegraphics[scale=0.9, clip=]{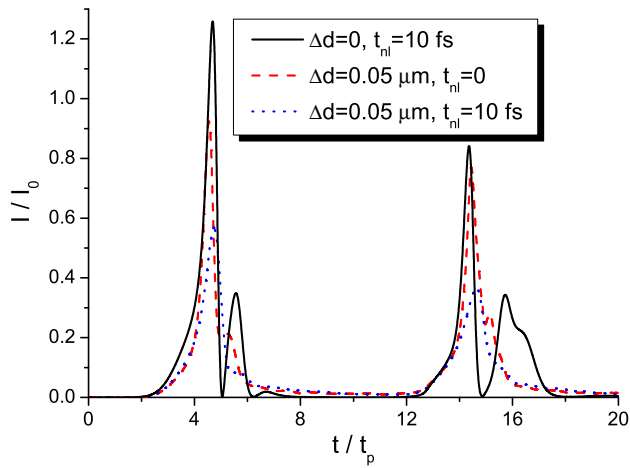}
\centering \caption{\label{fig3} (Color online) The profiles of
co-propagating pulses transmitted through the ordered and disordered
photonic crystals with and without nonlinearity relaxation. The
nonlinearity coefficient is $n_2 I_0=0.05$. The curves are averaged
over $50$ realizations.}
\end{figure}

\begin{figure}[t!]
\includegraphics[scale=0.84, clip=]{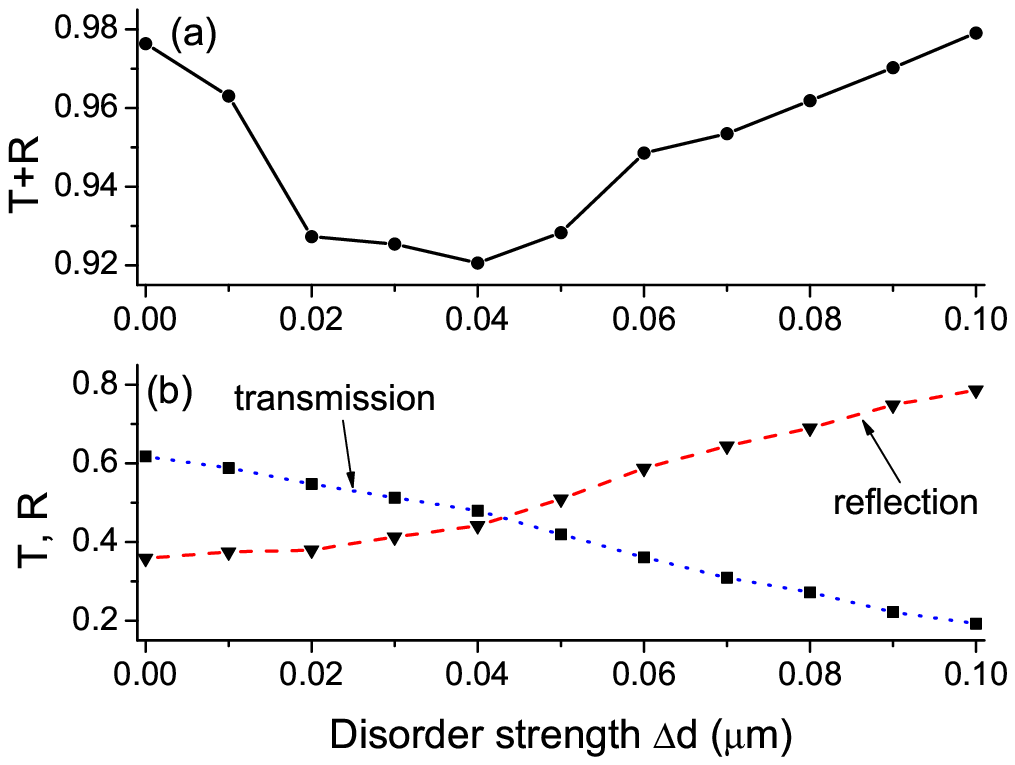}
\centering \caption{\label{fig4} (Color online) The dependencies of
(a) the average total output and (b) the average transmission and
reflection on the strength of disorder. The nonlinearity coefficient
is $n_2 I_0=0.05$, the relaxation time is $t_{nl}=10$ fs. The
averaging was made over $100$ realizations.}
\end{figure}

\begin{figure}[t!]
\includegraphics[scale=0.9, clip=]{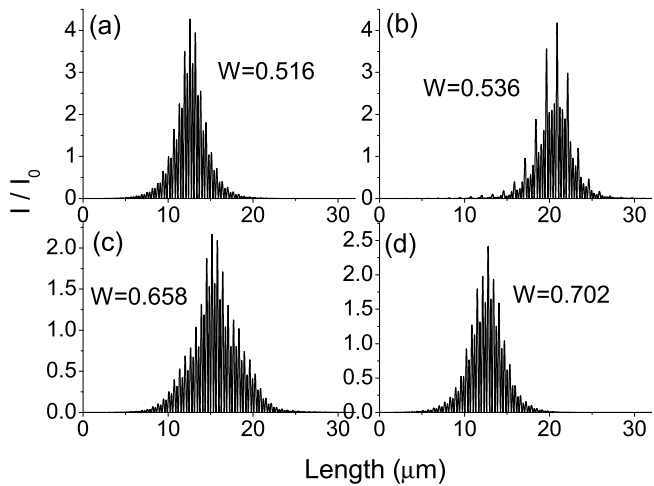}
\centering \caption{\label{fig5} The intensity distributions (at
$t=99t_p$) for several realizations of pulses co-propagating in the
structure with the nonlinearity coefficient $n_2 I_0=0.05$,
relaxation time $t_{nl}=10$ fs, and the disorder strength $\Delta
d=0.05$ $\mu$m.}
\end{figure}

\begin{figure}[t!]
\includegraphics[scale=0.88, clip=]{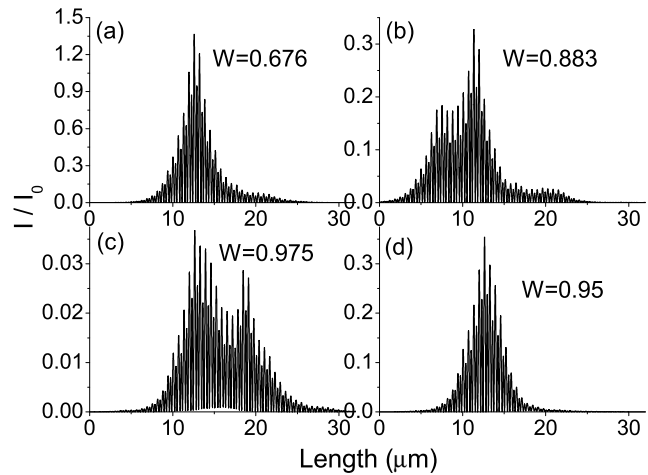}
\centering \caption{\label{fig6} (a)-(c) The intensity distributions
for several realizations of single pulse propagating in the
disordered structure with the same parameters as in Fig. \ref{fig5}.
Panels (a)-(b) correspond to the same realizations as in Fig.
\ref{fig5}(a) and Fig. \ref{fig5}(b), panel (d) shows the two-pulse
distribution for the realization of panel (c). }
\end{figure}

\begin{table*}[!]
  \caption{\label{tabcoprop}The data on realizations in the case of two pulses
  co-propagating through the disordered photonic crystal. The parameters $n_2 I_0=0.05$ and $t_{nl}=10$ fs are used,
  if the other is not stated. The simulation time is $100 t_p$.
  (The notations are used as follows: $\bar T$ is the average transmission, $\bar R$ the average reflection, $\bar W$ the average output energy, i.e. $\bar T + \bar R$;
  $N_{90}$ the number of realizations with the output energy $W > 0.9$, $N_{80}$ the number of realizations with $0.8< W < 0.9$,
  $N_{70}$ the number of realizations with $0.7< W < 0.8$, $N_{60}$ the number of realizations with $0.6< W < 0.7$,
  $N_{50}$ the number of realizations with $W < 0.6$; $N_t$ the total number of realizations; $W_{min}$ the minimal output energy $W$ among the realizations.)}
  \begin{ruledtabular}
    \begin{tabular}{ccccccccccc}
    \hline
    $\Delta d$ & $\bar T$ & $\bar R$ & $\bar W$ & $N_{90}$ & $N_{80}$ & $N_{70}$
    & $N_{60}$ & $N_{50}$ & $N_t$ & $W_{min}$\\
    \hline
    0 & 0.618 & 0.359 & 0.976 & 1 & - & - & - & - & 1 & 0.976\\
    0.01 & 0.588 & 0.375 & 0.963 & 94 & 6 & - & - & - & 100 & 0.809\\
    0.02 & 0.548 & 0.379 & 0.927 & 77 & 14 & 6 & 2 & 1 & 100 & 0.558\\
    0.03 & 0.512 & 0.413 & 0.925 & 79 & 7 & 8 & 5 & 1 & 100 & 0.527\\
    0.04 & 0.479 & 0.441 & 0.920 & 74 & 14 & 8 & 2 & 2 & 100 & 0.481\\
    0.05 & 0.419 & 0.509 & 0.928 & 82 & 4 & 7 & 5 & 2 & 100 & 0.516\\
    0.06 & 0.361 & 0.587 & 0.948 & 82 & 13 & - & 5 & - & 100 & 0.625\\
    0.07 & 0.309 & 0.644 & 0.953 & 83 & 13 & 3 & 1 & - & 100 & 0.653\\
    0.08 & 0.272 & 0.690 & 0.962 & 89 & 7 & 1 & 3 & - & 100 & 0.644\\
    0.09 & 0.222 & 0.748 & 0.970 & 89 & 11 & - & - & - & 100 & 0.824\\
    0.10 & 0.192 & 0.787 & 0.979 & 96 & 2 & 2 & - & - & 100 & 0.793\\
    0.05 (single pulse) & 0.450 & 0.521 & 0.971 & 93 & 5 & 1 & 1 & - & 100 & 0.649\\
    0.05 ($n_2 I_0=0.01$) & 0.45 & 0.5489 & 0.9989 & 50 & - & - & - & - & 50 & 0.9975\\
    0.05 ($t_{nl}=0$) & 0.5136 & 0.4782 & 0.9918 & 50 & - & - & - & - & 50 & 0.956\\
    \hline
    \end{tabular}
    \end{ruledtabular}
\end{table*}

First, let us consider the situation of two co-propagating pulses
launched into the structure with some interval one after another.
This interval must be not too large for the pulses to interact
effectively with each other and not too small so that we can talk
about separate pulses. In our calculations, we assume the interval
of $10 t_p$ between the peaks of the incident pulses. We start with
the profiles of the pulses transmitted through the perfect (ordered)
photonic crystal (Fig. \ref{fig1}). It is seen that the pulses have
different peaks even in the linear case. This means that the
interpulse interval is short enough to provide effective energy
interchange between them. Perhaps, in the linear case, some residual
radiation of the first pulse joins the second one, so that its
intensity grows. This simple picture is not applicable for the more
complicated nonlinear case. In nonlinear structure, the first pulse
is stronger compressed (more intense) than the second one. Figure
\ref{fig2} shows the changes in the profiles due to disorder with
$\Delta d=0.05$ $\mu$m. In the linear case, the averaged transmitted
pulses seem to be almost identical, i.e. on average, the
distribution of energy between the pulses is uniform. This
uniformity is broken as a result of nonlinearity introduction: the
first pulse tends to be more powerful than the second one. Now we
can add the relaxation of nonlinearity and study its influence on
the averaged profiles of the co-propagating pulses (Fig.
\ref{fig3}). It is seen that addition of relaxation to the
disordered structure results in further decrease of the intensity of
transmitted pulses.

What is the reason for this decrease? Does it mean simply
strengthening of reflection? The detailed study shows that the
answer is ``no''. According to the data shown in Table
\ref{tabcoprop}, the average transmission $\bar T$ (the part of
total light energy transmitted through the structure in the time
$100 t_p$ and averaged over realizations) drops due to the
relaxation from $0.514$ to $0.419$ (remind that we consider the
disorder strength $\Delta d=0.05$ $\mu$m). At the same time, the
reflection $\bar R$ averaged over realizations grows from $0.478$
only to $0.509$. This means that the total average output $\bar W$
(sum of transmission and reflection) decreases from almost unity to
$0.928$, i.e. \textit{on average} more than $7 \%$ of the input
energy remains inside the structure due to the relaxation of
nonlinearity. We further explored how the average output energy
depends on the disorder strength. The resulting curves presented in
Fig. \ref{fig4} show that, as it would be expected of the disordered
media, the transmission decreases and reflection increases with the
growing $\Delta d$. However, these two processes do not compensate
each other, so that the dip in the curve for the total output energy
appears. The minimum of $\bar W$ occurs at $\Delta d=0.04$ $\mu$m
and amounts to about $0.92$.

The data on average output implies that there may act the mechanism
analogous to the self-trapping effect reported in the previous
publications \cite{Novitsky2010, Novitsky2014}. This assumption is
justified by consideration of concrete realizations; the intensity
distributions (at the time instant $t=99 t_p$) along the structure
for some of realizations with comparatively low output $W$ are shown
in Fig. \ref{fig5}. These distributions correspond to the residual
radiation left in the disordered multilayer after passage of both
pulses. It is seen that the width and peak intensity of the
distributions strongly depends on the characteristics of the
concrete realization. For example, comparison of Fig. \ref{fig5}(a)
and \ref{fig5}(b) shows that, even at approximately the same value
of the output $W$, the distributions can strongly differ from each
other. As a rule, however, decrease in $W$ is accompanied by raise
of the peak intensity and by narrowing of the distribution. One can
compare these two distributions with the residual intensity
distributions after passage of a single pulse through the structure
with the same parameters (the same realization) shown in Fig.
\ref{fig6}(a) and \ref{fig6}(b). It is seen that the output is much
greater in the case of the single pulse, i.e. a significant part of
energy is trapped inside the system as a result of the interaction
of the pulses. The difference between the results in Fig.
\ref{fig6}(a) and \ref{fig6}(b) can be compared with the difference
of distributions in Fig. \ref{fig5}(a) and \ref{fig5}(b): in the
panel (b), light is trapped near the exit of the structure, so that
the light storage is not so stable as in the panel (a). Finally, in
Fig. \ref{fig6}(c) and \ref{fig6}(d), we plot the distributions of
single and double pulse residual intensities for the ``usual''
(large-output) realization. Though the difference in the absolute
value of $W$ is small, the distribution in the case of
co-propagating pulses has a characteristic symmetric shape implying
the formation of the (quasi)stable ``trap''.

Obviously, the examples discussed above can be treated as an
evidence of light trapping enhanced by interaction of two
co-propagating pulses. The statistical confirmation of this effect
is given in Table \ref{tabcoprop} which shows the number of
realizations with the outputs $W$ in the certain ranges for
different disorder strengths $\Delta d$. It is seen that, in
accordance with the data of Fig. \ref{fig4}, the number of
low-output realizations grows with increasing $\Delta d$, reaches
maximum at $\Delta d=0.04$ $\mu$m and then starts to decrease. At
$\Delta d=0.1$ $\mu$m, we have $W>0.9$ for almost all realizations.
The same fall and rise is characteristic for the value of minimal
output among the realizations at a given disorder strength (see the
last column of the table). The last three strings of Table
\ref{tabcoprop} allow us to compare the case of co-propagating
pulses with the cases of a single pulse, comparatively weak
nonlinearity and no relaxation at the same disorder (namely, $\Delta
d=0.05$ $\mu$m), respectively. This comparison shows that the
two-pulse scheme allows to strongly increase the efficiency of light
trapping inside the disordered photonic crystal.

Further, we have studied the propagation of a single pulse
containing the same energy as two interacting pulses considered
above, i.e. we dealt with the pulse of the peak intensity $2 I_0$.
Can the results on trapping enhanced by two interacting pulses be
compared with this single pulse case? Our calculations show that the
probability of high-intensity pulse trapping in disordered structure
with $\Delta d=0.05$ $\mu$m is much larger than in two-pulse scheme
of Table \ref{tabcoprop}: we have the average output $\bar W \approx
0.774$, $W_{min}=0.416$ and the number of realizations with $W<0.6$
as large as $21$ (from the total number of $100$). However, such
great efficiency of trapping has simple explanation: high-intensity
pulse trapping can be observed already in the ordered system giving
$W=0.640$. This is the fundamental difference with the situation
reported above for the pulses of lower intensities which can be
trapped only in the presence of disorder. For the high-intensity
pulse, the situation seems to be inverted: the disorder leads to
some degradation of trapping, since the average output $\bar W$ is
higher than the output for the ordered photonic crystal. Thus, if we
want to have the effect of disorder on interacting pulses discussed
in this section, the intensity of pulses should be not too high:
under this condition, the self-trapping can be observed neither for
the single pulse in the disordered structure nor for the
co-propagating pulses in the ordered structure. Further, we will
deal only with the pulses of appropriate intensity.

The results of calculations reported in this section make it clear
that there is an optimal level of disorder for observation of
trapping of energy of co-propagating pulses. This fact along with
the absence of trapping in the perfect (ordered) structure is the
reason for us to call this effect \textit{the disorder-induced light
trapping} in the photonic crystal.

\section{\label{counterprop}Counter-propagating pulses}

\begin{table*}[t!]
  \caption{\label{tabcounter}The data on realizations in the case of two pulses
  counter-propagating through the disordered photonic crystal. The parameters $n_2 I_0=0.05$ and $t_{nl}=10$ fs are used,
  if the other is not stated. The simulation time is $100 t_p$.}
  \begin{ruledtabular}
    \begin{tabular}{ccccccccccc}
    \hline
    $\Delta d$ & $\bar T$ & $\bar R$ & $\bar W$ & $N_{90}$ &
    $N_{80}$ & $N_{70}$ & $N_{60}$ & $N_{50}$ & $N_t$ & $W_{min}$\\
    \hline
    0 & 0.492 & 0.488 & 0.980 & 1 & - & - & - & - & 1 & 0.980\\
    0.01 & 0.494 & 0.482 & 0.976 & 100 & - & - & - & - & 100 & 0.952\\
    0.02 & 0.478 & 0.490 & 0.968 & 98 & 2 & - & - & - & 100 & 0.826\\
    0.03 & 0.480 & 0.474 & 0.954 & 92 & 4 & 1 & 2 & 1 & 100 & 0.578\\
    0.04 & 0.494 & 0.467 & 0.960 & 90 & 8 & 2 & - & - & 100 & 0.744\\
    0.05 & 0.491 & 0.460 & 0.951 & 87 & 7 & 4 & 1 & 1 & 100 & 0.594\\
    0.06 & 0.477 & 0.481 & 0.958 & 90 & 7 & 2 & 1 & - & 100 & 0.674\\
    0.07 & 0.487 & 0.475 & 0.962 & 89 & 8 & 3 & - & - & 100 & 0.748\\
    0.08 & 0.482 & 0.491 & 0.973 & 94 & 6 & - & - & - & 100 & 0.824\\
    0.09 & 0.505 & 0.468 & 0.973 & 98 & 1 & 1 & - & - & 100 & 0.743\\
    0.10 & 0.496 & 0.476 & 0.972 & 95 & 2 & 1 & 2 & - & 100 & 0.641\\
    0.05 (single pulse) & 0.450 & 0.521 & 0.971 & 93 & 5 & 1 & 1 & - & 100 & 0.649\\
    0.05 ($n_2 I_0=0.01$) & 0.484 & 0.515 & 0.999 & 50 & - & - & - & - & 50 & 0.997\\
    0.05 ($t_{nl}=0$) & 0.500 & 0.494 & 0.994 & 50 & - & - & - & - & 50 & 0.957\\
    \hline
    \end{tabular}
   \end{ruledtabular}
\end{table*}

\begin{figure}[t!]
\includegraphics[scale=0.82, clip=]{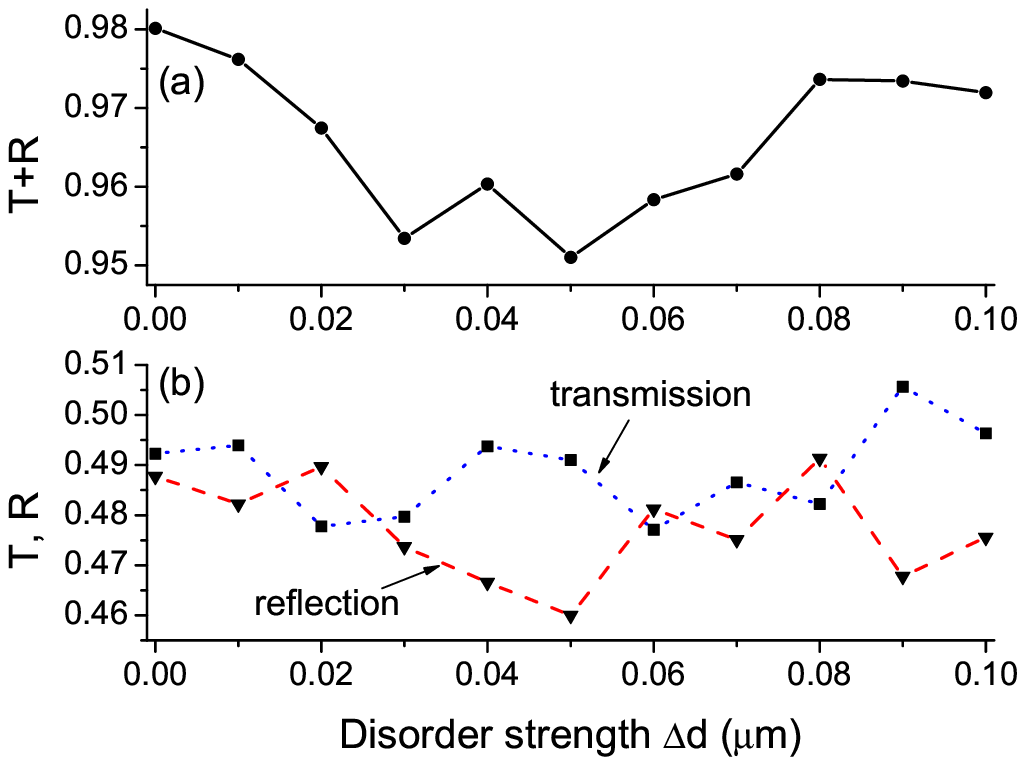}
\centering \caption{\label{fig7} (Color online) The dependencies of
(a) the average total output and (b) the average transmission and
reflection on the strength of disorder in the case of
counter-propagating pulses. The nonlinearity coefficient is $n_2
I_0=0.05$, relaxation time $t_{nl}=10$ fs. The averaging was made
over $100$ realizations.}
\end{figure}

\begin{figure}[t!]
\includegraphics[scale=0.84, clip=]{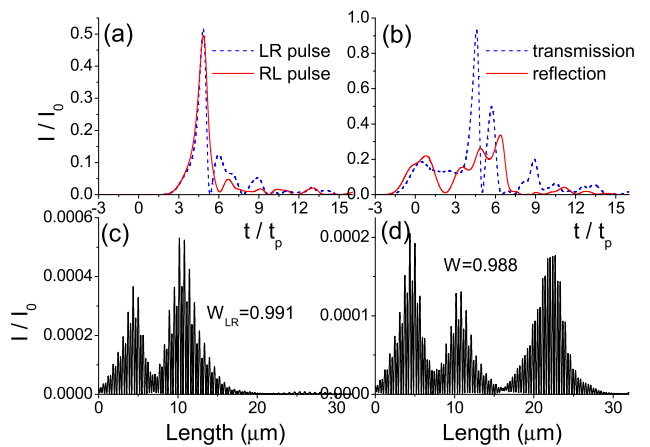}
\centering \caption{\label{fig8} The example of specific realization
with $W=0.988$ in counter-propagating regime. (a) Profiles of single
LR and RL pulses transmitted through the structure with a given
disorder, (b) profiles of transmitted and reflected light when both
pulses are launched into the structure, (c) and (d) intensity
distributions (at the time instant $t=99 t_p$) corresponding to the
situations of (a) and (b), respectively. The nonlinearity
coefficient is $n_2 I_0=0.05$, relaxation time $t_{nl}=10$ fs, and
the disorder strength $\Delta d=0.05$ $\mu$m.}
\end{figure}

\begin{figure}[t!]
\includegraphics[scale=0.9, clip=]{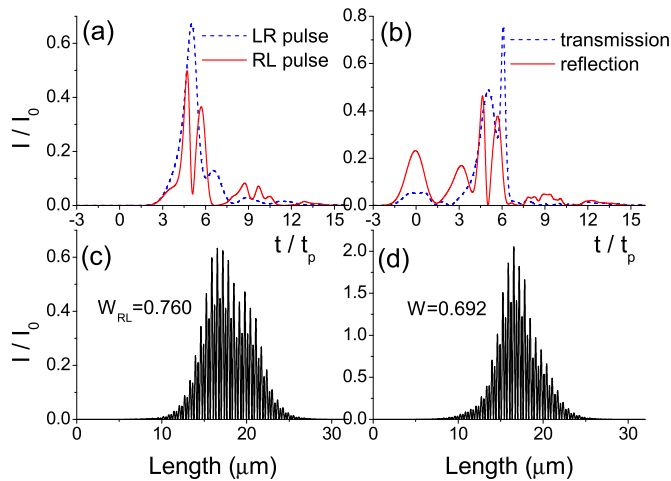}
\centering \caption{\label{fig9} The same as in Fig. \ref{fig8} but
for another realization with $W=0.692$ in counter-propagating
regime.}
\end{figure}

In this section, we consider another situation when the interacting
pulses are counter-propagating. We restrict ourselves to the
symmetric scheme when both pulses are identical. This means that, on
average, reflection and transmission through the photonic crystal
should be equal, i.e. $\bar T=\bar R \approx 0.5$. Here the
convention is used as follows: We call "reflection" the part of
total energy outgoing from, say, the left edge of the structure,
i.e. it is generated by the reflected light of the pulse propagating
from left to right (LR) and transmitted light of the pulse
propagating from right to left (RL); hence, for "transmission" we
have the energy part from the right edge, i.e. transmission of LR
pulse plus reflection of RL pulse.

Is it possible to observe the disorder-induced trapping as a result
of collision of counter-propagating pulses? The answer is not
obvious, since the time of interaction in this scheme seems to be
much shorter than for co-propagating pulses moving side by side
along the whole length of the structure. We performed calculations
for different disorder strengths in the manner of previous section.
The statistics of realizations is represented in Table
\ref{tabcounter}, while Fig. \ref{fig7} shows the average values of
output energy. It is seen that transmission and reflection vary
around the same average level, but the total output $\bar W$ and the
minimal output energy $W_min$ have the lowest values in the range
$\Delta d=0.03 - 0.05$ $\mu$m. This is in accordance with the
results of the previous section [see Fig. \ref{fig4}(a)], though the
maximal part of energy remaining inside the photonic crystal drops
from about $7 \%$ (for co-propagating pulses) to less than $5 \%$
(for counter-propagating pulses). The value of the disorder strength
at which the output minimum occurs ($\Delta d=0.05$ $\mu$m) is,
perhaps, connected with the parameters of the structures considered,
in particular with the average thicknesses of the layers: the
amplitude of disorder should be high enough comparing with these
thicknesses to observe influence of disorder, but not too high, so
that the structure can still be considered as periodic on average.

The appearance of disorder-induced trapping is confirmed by the
statistics of realizations shown in Table \ref{tabcounter}: the
number of realizations with low output first increases with disorder
strength and then decreases. The same is true for the minimal output
$W_{min}$ among the realizations. The last three strings of the
table calculated for control corroborate the importance of
collisions, high enough nonlinearity and presence of relaxation to
obtain effective trapping. The effect is less pronounced in
comparison with the case of co-propagating pulses, perhaps, because
of lower interaction time as mentioned above.

Let us consider two typical realizations. One of them shown in Fig.
\ref{fig8} does not reveal any substantial trapping due to the
collision (the output energy is $W=0.988$). Another realization
demonstrated in Fig. \ref{fig9} is characterized by trapping of
approximately $30 \%$ of energy of the colliding pulses. If there is
only one pulse (either LR or RL), then transmission is almost
identical for both directions of propagation in the first case [see
Fig. \ref{fig8}(a)]. The collision generally breaks this symmetry,
so that transmission and reflection (in the sense discussed above)
have very different intensity profiles [Fig. \ref{fig8}(b)]. On the
contrary, in the strong trapping regime, there is no transmission
symmetry even in the absence of counter-propagating pulse [Fig.
\ref{fig9}(a)]. Finally, we should consider the distributions of
residual light intensity along the structure [panels (c) and (d)].
In the realization shown in Fig. \ref{fig8}, collision does not
qualitatively change the distribution: there is still several peaks
of very low intensity. Fundamentally another situation is seen in
Fig. \ref{fig9}: the collision results in formation of a single
high-intensity bell-shaped peak which can with every reason be
called ``the trap''. Figure \ref{fig9}(c) shows the distribution
only for the RL pulse with $W_{RL}=0.76$, while for LR pulse we have
$W_{RL}=0.993$, so that in this last instance there is no any
substantial light trapping. The trap becomes more intensive and
symmetric in shape (and, hence, more stable) in the case of
colliding pulses [Fig. \ref{fig9}(d)]. Of course, there is
possibility that the collision will make trapping less effective,
but, according to the statistics discussed above, the inverse
situation is more probable.

\section{Conclusion}

In this paper, we predict the possibility of observing light
trapping enhanced by collisions of pulses in disordered photonic
crystals with relaxing cubic nonlinearity. Since there is an optimal
value of disorder, we call this effect the disorder-induced
trapping. At very low (or no) disorder strengths and at very high
disorder strengths, the probability of effective light trapping
(measured as a number of realizations with low output energy) is
strongly suppressed. It is also necessary to have high enough
nonlinearity coefficients with nonzero relaxation times, but not too
high so that the purely nonlinear trapping to be absent in the
ordered case. Though, at these conditions, some part of energy can
be trapped even in the single-pulse regime, interaction of pulses
(either co-propagating or counter-propagating) strongly increase the
number of realizations with effective trapping. Our observations can
be considered as the preliminary report on the possibility of this
effect. More investigations are needed to search for the optimal
parameters or to study the effect with larger number of interacting
pulses.

\acknowledgements{The work was supported by the Belarusian State
Foundation for Fundamental Research, project F13M-038.}

\end{document}